\documentclass[aps,prl,twocolumn,superscriptaddress]{revtex4}

\usepackage{graphicx}

\begin{document}


\title{Direct evaporative cooling of $^{41}$K into a Bose-Einstein condensate
}


\author{T. Kishimoto}
\affiliation{University of Electro-Communications, 1-5-1 Chofugaoka, Chofu 182-8585, Japan}
\author{J. Kobayashi}
\affiliation{Institute of Engineering Innovation, University of Tokyo, Bunkyo-ku, Tokyo 113-8656, Japan}
\author{K. Noda}
\affiliation{Department of Applied Physics, University of Tokyo, Bunkyo-ku, Tokyo 113-8656, Japan}
\author{K. Aikawa}
\affiliation{Department of Applied Physics, University of Tokyo, Bunkyo-ku, Tokyo 113-8656, Japan}
\author{M. Ueda}
\affiliation{Department of Physics, University of Tokyo, Hongo, Bunkyo-ku, Tokyo 113-0033, Japan}
\affiliation{ERATO Macroscopic Quantum Control Project, JST, Bunkyo-ku, Tokyo 113-8656, Japan}
\author{S. Inouye}
\email{inouye@sogo.t.u-tokyo.ac.jp}
\affiliation{Institute of Engineering Innovation, University of Tokyo, Bunkyo-ku, Tokyo 113-8656, Japan}
\affiliation{ERATO Macroscopic Quantum Control Project, JST, Bunkyo-ku, Tokyo 113-8656, Japan}

\date{\today}

\begin{abstract}
We have investigated the collisional properties of ${}^{41}$K atoms at ultracold temperature.
To show the possibility to use ${}^{41}$K as a coolant,
a Bose-Einstein condensate of ${}^{41}$K atoms in the stretched state ($F=2, m_F=2$) was created for the first time by direct evaporation in
a magnetic trap.
An upper bound of three body loss coefficient for atoms in the condensate was determined to be $4(2) \times 10^{-29} {\rm cm}^{-6}{\rm s}^{-1}$.
A Feshbach resonance in the $F=1, m_F=-1$ state was observed
at 51.42(5) G, which is in good agreement with theoretical prediction.
\end{abstract}


\maketitle

Recently, a dilute gas quantum degenerate Bose-Fermi mixture is attracting a lot of
interest. Mean-field-driven collapse was observed with
${}^{40}$K and ${}^{87}$Rb in a harmonic trap\cite{modu02science,cospe06}.
Collisionally induced transport was observed with a Bose-Fermi mixture in a 1D lattice\cite{ott04}.
There has been a significant advance in understanding the behavior
of the mixture in a three dimensional optical lattice\cite{gunt06}.
Also, a variety of new phases were predicted for Bose-Fermi mixtures in optical lattices\cite{lewe04}.
Furthermore, quest for Boson-induced Fermion superfluidity has been considered as
one of the best testing grounds for simulating the real superconductivity in solids\cite{boson-mediated}.
To date, a quantum degenerate Bose-Fermi mixture has been reported for
${}^{6}$Li-${}^{7}$Li\cite{trus01},
${}^{6}$Li-${}^{23}$Na\cite{hadz02},
${}^{40}$K-${}^{87}$Rb\cite{roat02},
${}^{6}$Li-${}^{87}$Rb\cite{silb05},
${}^{3}$He-${}^{4}$He\cite{mcna06},
${}^{6}$Li-${}^{40}$K-${}^{87}$Rb\cite{tagl08}, and
${}^{173}$Yb-${}^{174}$Yb\cite{fuku08},
where ${}^{3}$He,${}^{6}$Li,${}^{40}$K, and ${}^{173}$Yb are fermions.

Often, manipulation of a trapped hetero-nuclear ultracold dilute gas mixture needs
extra attention due to difference in gravitational sags.
Inhomogeneous spatial overlap between the two clouds makes the analysis less intuitive.
Popular solutions for this problem are to increase the vertical trapping
frequencies, to make the trapping potential sensitive to atomic species\cite{ospe08},
or to use atomic species close in mass. The last solution favors
``isotopic Bose-Fermi mixture,'' since the mass difference is minimized.
This scheme has been realized for He, Li and Yb, but not for K.

Here we report our study on collisional properties of ${}^{41}$K.
Our main motivation was to establish ${}^{41}$K as a new coolant,
enabling the production of a new isotopic Bose-Fermi mixture of ${}^{41}$K-${}^{40}$K.
Starting from a pure ${}^{41}$K sample, we have produced a pure Bose-Einstein condensate
with more than $3 \times 10^5$ atoms. Evaporation trajectory resembles
quite remarkably to ${}^{87}$Rb. Indeed, we measured the efficiency of
evaporative cooling of both pure ${}^{41}$K sample and pure ${}^{87}$Rb sample in the same environment
and found they were only different by  about $15$\%.

\begin{figure}
\begin{center}
\includegraphics[height=70mm]{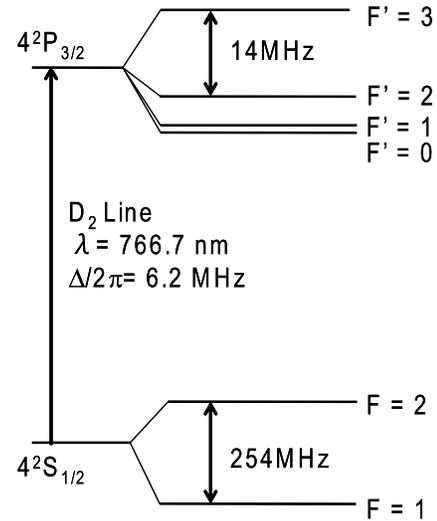}
\end{center}
\caption{\label{fig:41K_level_scheme} Level diagram of the $4\,{}^{2}{S}_{1/2}$ state and the $4\,{}^{2}{P}_{3/2}$ state of ${}^{41}$K atom.
The hyperfine splitting of the $4\,{}^{2}{P}_{3/2}$ state is comparable to the natural
line width of the $D_2$ line (6.2MHz).}
\end{figure}

Potassium has another stable bosonic isotope (${}^{39}$K), and it is more abundant (93.3\%) compared to ${}^{41}$K(6.7\%).
We chose ${}^{41}$K,
just because the triplet scattering length of ${}^{39}$K is negative,
and Feshbach resonance is needed for producing a large condensate\cite{roat07}.
To date, most of the collisional properties of ${}^{41}$K have been inferred from measurements done
on other isotopes, rather than explored directly using ${}^{41}$K.
The current best estimate for the triplet scattering length of ${}^{41}$K is
$a_{41}\sim 60\,a_0$ where $a_0$ is the Bohr radius\cite{erri07}.
This value was obtained by combining the results from two-photon spectroscopy of ${}^{39}$K\cite{wang00} and Feshbach spectroscopy
of ${}^{39}$K\cite{erri07}, ${}^{40}$K\cite{loft02}.
${}^{41}$K has been Bose condensed using Rb as a coolant\cite{modu01science}.

For investigating the collisional properties of ${}^{41}$K experimentally, we followed the general method
for producing quantum degenerate gas in a trap, namely laser cooling followed by evaporative cooling in
a magnetic trap. Laser cooling of ${}^{41}$K needs extra attention.
The peculiar feature of ${}^{41}$K is that the hyperfine splittings are small. Especially, the hyperfine splitting
in the $4\,{}^{2}{P}_{3/2}$ state is only $\sim 14$ MHz, which is on the same order
as the natural linewidth (6.2 MHz). Thus, it is practically impossible to
form a closed transition, since off-resonant excitation to other excited state is not at all
negligible. Indeed, in order to increase the capture range of the MOT, we had to
detune the ``cooling beam'' to the red of the whole excited state hyperfine manifold\cite{raman}.
By preparing almost the same amount of laser power for both ``cooling'' and ``repumping'' transitions,
we have overcome these problems and succeeded in capturing a large
number of ${}^{41}$K atom in the MOT\cite{cata06}.

Details of our experimental setup can be seen elsewhere\cite{kish08}.
We employ a double-MOT system.
The first MOT in the first glass cell was loaded from the background gas, which was released from a homemade dispenser\cite{dema99}.
The dispenser has 98.7\% ${}^{41}$K enriched KCl. The first cell was kept at 50 ${}^{\circ}$C in order to keep the loading rate of the first MOT.
A beam of atoms were pushed from the first MOT to the second MOT using a resonant push beam.
The frequency of the push beam is 6MHz red-detuned from $4 S_{1/2}$ $F=1$ to $4 P_{3/2}$ $F'=2$ transition.
We load $ 1.4 \times 10^9 $ ${}^{41}$K atoms in the 2nd MOT in 10 seconds.
Since typical temperature of atoms in ${}^{41}$K MOT was $\sim 5$mK,
we introduced additional stages for spatial compression(40ms) and Doppler cooling(12ms)\cite{aikawa}.
Typical number, density, and temperature of the cloud after the cooling are $1.0 \times 10^{9}$,
$5.1 \times 10^{10}{\rm cm}^{-3}$, and $100\,\mu$K, respectively.
The obtained phase-space density is $\sim 1.0 \times 10^{-6}$, which is adequate for starting evaporative cooling.

\begin{figure}
\includegraphics[width=80mm]{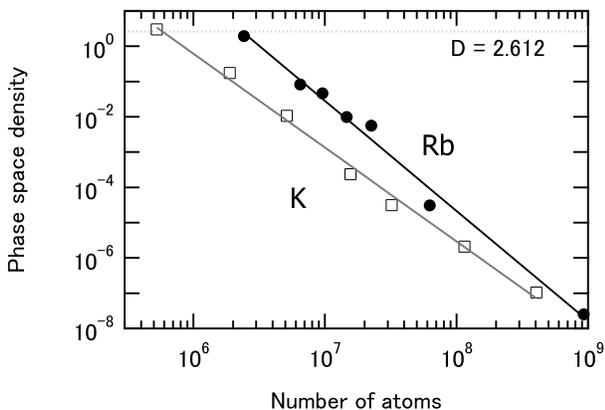}%
\caption{\label{fig:evaporation} Typical evaporative cooling trajectory for ${}^{41}$K (square) and ${}^{87}$Rb (circle) atoms.
The efficiency of evaporative cooling $\gamma = -d(\ln D)/d(\ln N)$ was 2.7 for ${}^{41}$K and 3.1 for ${}^{87}$Rb.
Here, $D$ is the phase-space density of the gas.
}
\end{figure}

\begin{table}
\caption{\label{table:evaporation}A comparison between typical conditions before and after the
evaporative cooling for ${}^{41}$K and ${}^{87}$Rb.}
\begin{ruledtabular}
\begin{tabular}{cccc}
species & initial PSD & $N_{\rm BEC}$ & $\gamma$\\
${}^{41}$K & $4.5\times10^{-7}$ & $3 \times 10^5$& 2.7\\
${}^{87}$Rb & $2.5\times10^{-8}$ & $7 \times 10^5$& 3.1\\

\end{tabular}
\end{ruledtabular}
\end{table}

The efficiency of evaporative cooling is difficult to predict, since it is quite
sensitive to small inelastic loss processes. In general, the efficiency
of evaporative cooling depends sensitively on the ratio between ``good''(elastic) and ``bad''
(inelastic) collisions, which is typically more than factor of a few hundred for alkali atoms.
Evaporative cooling of alkali atoms in the stretched state is usually
quite efficient, since one can expect a very tight confinement from the trap,
and there should be no scattering resonances. A prominent
exception has been ${}^{133}$Cs, whose evaporative cooling in the stretched
state was hindered by the strong dipolar relaxation process induced by the second-order
spin-orbit interaction\cite{leo98}.

For loading ${}^{41}$K atoms into a magnetic trap, we optically pumped
the atoms into the stretched  state($|F,m_F\rangle=|2,2\rangle$).
The weak-field seeking state in the lower hyperfine manifold ($|1,-1\rangle$) is not
ideal for magnetic trapping, since the maximum trap depth for $|1,-1\rangle$ state is
limited to $\sim 0.8\,{\rm mK}$ due to the Paschen-Back effect.
Our magnetic trap is in the Ioffe-Pritchard configuration.
In order to minimize the heating during the transfer to the magnetic trap,
we load atoms in a loose magnetic trap with high bias field,
and then atoms are adiabatically compressed by increasing the radial confinement and lowering the bias field.
The trapping frequency after compression is $\nu_r=325$ Hz ($\nu_z=15.0$ Hz) in the radial (axial) direction.
The typical initial condition after adiabatic compression is $N=8.2\times 10^8$ and $T \sim 680\,\mu{\rm K}$,
which gives the phase-space density of $4.5 \times 10^{-7}$.

Figure \ref{fig:evaporation} shows a typical evaporation trajectory
for ${}^{41}$K atoms in the $|F,m_F\rangle=|2,2\rangle$ state.
We first tried to evaporatively cool ${}^{41}$K atoms by driving the rf-transition between Zeeman sublevels.
However, we found it was necessary to actively remove atoms in the
$|2,1\rangle$ and $|2,0\rangle$ states. We think this purification was necessary since a large 2nd-order
Zeeman shift increases the accumulation of atoms in the $|2,1\rangle$ and $|2,0\rangle$
states, and also increases the energy release from $|2,1\rangle+|2,1\rangle \rightarrow |2,2\rangle+|2,0\rangle$
collisions\cite{hadz03,desr99}. For obtaining a pure $|2,2\rangle$ sample,
we did two things: we selectively removed atoms in the $|2,1\rangle$ and $|2,0\rangle$ states
by driving rf transitions to the lower hyperfine manifold when the bias field is high.
We also decided to evaporate the cloud by driving the $|2,2\rangle$ $\rightarrow$ $|1,1\rangle$ transition, which is 254 MHz at zero magnetic field.

The efficiency of evaporative cooling is defined as $\gamma=-d(\ln D)/d(\ln N)$, where $D$ is the phase-space
density of the gas\cite{kett96}.
We observed efficiency of $\gamma_{\rm K}=2.7$ in our system.
This is comparable with or even better than some of the typical $\gamma$ values for other alkali gases.
For example, alkali-gas $\gamma$ values shown in ``Overview of Evaporative Cooling Experiments'' in Ref.\cite{kett96}
range from 1.5 to 2, with a prominent exception of ${}^{87}$Rb which shows $\gamma=3.0$.
Indeed, by completing the rf-sweep for evaporation, one can pass the phase-transition
for Bose condensation of ${}^{41}$K and obtain a pure condensate.
A typical number of atoms in the pure condensate was $3 \times 10^5$.

\begin{figure}
\includegraphics[width=85mm]{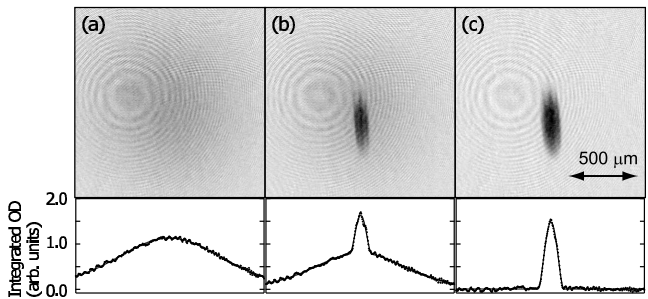}%
\caption{\label{fig:41KBEC}
A Bose-Einstein condensate of ${}^{41}$K atoms obtained by direct evaporation in a magnetic trap.
Absorption images of ${}^{41}$K atoms after 40 ms expansion show
(a) thermal cloud at T=750 nK, (b)bimodal distribution at T=580 nK, and (c)almost pure
condensate of $3 \times 10^5$ atoms.
The evaporative cooling was performed by driving rf-transition between hyperfine levels of ${}^{41}$K atoms.
No other coolant was involved in the experiment.}
\end{figure}

For a more direct comparison, we prepared a gas of ${}^{87}$Rb atoms in the $|F,m_F\rangle=|2,2\rangle$ state
and did rf-induced evaporative cooling in the same magnetic trap\cite{Rb_evap}.
A typical evaporative cooling trajectory is shown in filled circles in Fig.\ \ref{fig:evaporation}.
The initial condition was $N_{\rm Rb}=9.3\times 10^8$ and $T_{\rm Rb} \sim 1.2\,{\rm mK}$.
The number of atoms in the pure ${}^{87}$Rb condensate was $7 \times 10^5$.
The obtained efficiency was $\gamma_{\rm Rb}=3.1$, which was only $\sim 15\%$ higher than ${}^{41}$K.
This difference in efficiency can be partly attributed to the difference in the elastic cross section.
The cross-section for ${}^{87}$Rb is about a factor of 2.8 larger than the one for ${}^{41}$K.

The three-body loss coefficient is important, since it practically limits the highest density one can achieve with a condensate.
By studying the decay of the condensate, we determined the upper bound for the three-body loss coefficient for ${}^{41}$K atoms
in the $|F,m_F\rangle = |2,2\rangle$ state. We have taken a decay curve for number of atoms in a condensate,
and fitted the data assuming exponential decay accelerated by the three-body loss processes.
The obtained three-body loss coefficient was $4(2) \times 10^{-29}{\rm cm}^{-6}{\rm s}^{-1}$.
We have to point out that this value is just an upper bound for the three-body loss, since we could not rule out
any contributions from the two-body loss processes.
This coefficient was slightly larger than  ${}^{23}{\rm Na}(1.6\times 10^{-29}{\rm cm}^{-6}{\rm s}^{-1}$)\cite{gorl03} or
${}^{87}{\rm Rb}(1.8\times 10^{-29}{\rm cm}^{-6}{\rm s}^{-1}$)\cite{sodi99}  in the upper hyperfine state.

\begin{figure}
\includegraphics[width=75mm]{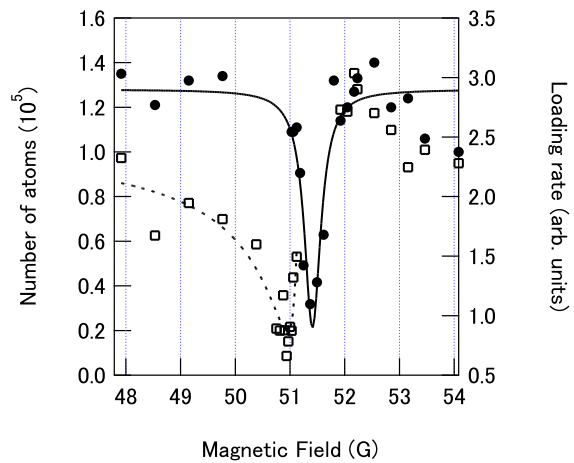}%
\caption{\label{fig:41K_Feshbach} A Feshbach resonance of ${}^{41}$K in the $|F,m_F\rangle=|1,-1\rangle$ state.
The number of trapped atoms (circle) shows a pronounced dip at a Feshbach resonance due to an enhanced three-body loss.
The resonance position was extracted from the data by fitting the loss with a Lorenzian(solid line).
The rate of collision-assisted loading (square) was used to locate the minimum of the elastic cross-section.
Assuming a resonant feature in the loading rate\cite{koba_fit}, we determined the width of the resonance.
The extracted center and the width of the resonance were 51.42(5) G and -0.47(5) G, and
they were in good agreement with theoretical predictions of Ref.\ \cite{erri07} (51.4 G and -0.3 G, respectively).}
\end{figure}

We also performed ``Feshbach spectroscopy'' for ${}^{41}$K atoms in the $F=1$ state.
By determining the position of Feshbach resonances, it is possible to check the validity of assumptions used for calculation.
A 300 mG wide Feshbach resonance is predicted for ${}^{41}$K atoms in the $|F,m_F\rangle=|1,-1\rangle$ state at 51.4 G\cite{erri07}.
That is the only $s$-wave Feshbach resonance predicted for ${}^{41}$K atoms in the $F=1$ manifold below 100 G.
Here, only resonances wider than 1 mG are considered.

We loaded the ${}^{41}$K atoms into a crossed-beam optical dipole
trap, which was produced by crossing two laser beams whose wavelength is 820 nm.
We measured the trap loss rate as a function of the bias magnetic field.
A pronounced loss feature was observed for atoms in the $|1,-1\rangle$ state near 51.4 G.
By assuming symmetric loss feature around the resonance, the center was determined to be 51.42(5)G,
which is in good agreement with theory(51.4 G).

The width of the resonance was determined by identifying the zero-crossing of the scattering length.
Minimum of elastic cross section was identified by monitoring how efficiently atoms were accumulated into the crossed-beam optical dipole trap.
Atoms were first loaded to a single beam optical dipole trap.
Then the second beam was switched on suddenly while the bias field was on.
The rate of atoms loaded to the crossed-region was plotted as a function of bias magnetic field
(figure \ref{fig:41K_Feshbach}). The minimum was observed at 50.95(5) G.
The observed width of the Feshbach resonance is -0.47(5) G, which is in fair agreement with
theory(-0.3 G).

In conclusion, we studied the collisional properties of ${}^{41}$K atoms.
A Bose-Einstein condensates of ${}^{41}$K atoms in the $|F,m_F\rangle=|2,2\rangle$ state was created by direct evaporation in
a magnetic trap. The efficiency of evaporative cooling was $\gamma=2.7$, which was only $\sim 15$\% lower than ${}^{87}$Rb under the same condition.
By measuring the lifetime of a condensate,
an upper bound of the three body loss coefficient was determined to be $4(2) \times 10^{-29} {\rm cm}^{-6}{\rm s}^{-1}$.
A Feshbach resonance in the $|F,m_F\rangle=|1,-1\rangle$ state was observed
at 51.42(5) G which was in good agreement with theory.
The observed high efficiency of evaporative cooling makes ${}^{41}$K a good candidate for a coolant.
Also we can expect a good thermalization between ${}^{40}$K and ${}^{41}$K since the interspecies scattering
length is relatively large and positive($a_{40-41}=97 a_0$)\cite{wang00}.
This opens up the possibility to perform Bose-Fermi mixture experiments using a new mixture ${}^{41}$K and ${}^{40}$K,
which should have negligible difference in sag due to gravity.
Even in general, ${}^{41}$K can be a good substitute for ${}^{87}$Rb. This is especially true when one wants to cool
atomic species that has a poor elastic cross section with ${}^{87}$Rb\cite{silb05}.

K.A. acknowledges support from the Japan Society for the Promotion of Science.


\begin{thebibliography}{99}
\bibitem{raman}
Since ``repumping beam'' is also detuned to the red of the whole excited state
manifold, the level scheme resembles more to a $\lambda$-type three-level system,
rather than the conventional four-level system for a magneto-optical trap for alkali gases.
We saw a small increase in number and a large increase in temperature of atoms in the MOT when we accidentally set
the two-photon detuning of the $\lambda$ system to zero.
\bibitem{aikawa} Spatial compression was achieved by increasing the magnetic field gradient
and reducing the repumping beam intensity.
For Doppler cooling, we reduced the detuning of the cooling beam and the intensity
of the repump beam.
\bibitem{Rb_evap} Rf-transitions between Zeeman sublevels were used for the evaporation.
Since we used the same magnetic field gradient for trapping the both species, the trapping frequencies are
different by a factor of square root of the mass ratio.
\bibitem{koba_fit}
The fitting function assumes there is a small offset in the loading rate.
This offset can originate from either trap unharmonicities or magnetic field fluctuations.
\end{thebibliography}
\end{document}